\documentclass[10pt]{iopart}

\usepackage{graphicx}
\usepackage{iopams}  
\usepackage{multirow}  
\usepackage{cite}

\begin{document}

\title[Blume-Capel model on nonregular lattices]{Monte Carlo studies of the Blume-Capel model on nonregular two- and three-dimensional lattices: Phase diagrams, tricriticality, and critical exponents
}

\author{Mouhcine Azhari}
\address{Fakult\"at f\"ur Mathematik und Naturwissenschaften Bergische Universit\"at Wuppertal, 42097 Wuppertal, Germany}
\ead{azhari.mouhcine@gmail.com}

\author{Unjong Yu}
\address{Department of Physics and Photon Science \&
         Research Center for Photon Science Technology,
          Gwangju Institute of Science and Technology,\\ 
          Gwangju 61005, South Korea}
\ead{uyu@gist.ac.kr (corresponding author)}

\date{\today}

\begin{abstract}
We perform Monte Carlo simulations, combining both the Wang-Landau and the Metropolis algorithms, to investigate the phase diagrams of the Blume-Capel model on different types of nonregular lattices (Lieb lattice (LL), decorated triangular lattice (DTL), and decorated simple cubic lattice (DSC)). The nonregular character of the lattices induces a double transition (reentrant behavior) in the region of the phase diagram at which the nature of the phase transition changes from first-order to second-order. A physical mechanism underlying this reentrance is proposed. The large-scale Monte Carlo simulations are performed with the finite-size scaling analysis to compute the critical exponents and the critical Binder cumulant for three different values of the anisotropy $\Delta/J \in \big\{ 0, 1, 1.34 \textrm{ (for LL)}, 1.51 \textrm{ (for DTL and DSC)} \big\}$, showing thus no deviation from the standard Ising universality class in two and three dimensions. We report also the location of the tricritical point to considerable precision: ($\Delta_t/J=1.3457(1)$; $k_B T_t/J=0.309(2)$), ($\Delta_t/J=1.5766(1)$; $k_B T_t/J=0.481(2)$), and ($\Delta_t/J=1.5933(1)$; $k_B T_t/J=0.569(4)$) for LL, DTL, and DSC, respectively.
\end{abstract}



\section{Introduction}
Magnetic phase transitions continue to attract attention even at present in both theory~\cite{Nishimori11,Zinn10,McCoy14} and experiment~\cite{Binder86,BELANGER2000,Wildes17}. Throughout the years, studies of statistical models of phase transitions with more than two states have provided tremendous insights in understanding these transitions in diverse research areas, in and beyond magnetism, including $^\mathrm{3}$He-$^\mathrm{4}$He mixtures~\cite{Graf67,Goellner71} and metamagnets~\cite{Schmidt70}. The Blume-Capel (BC) model~\cite{Blume66,Capel66,CAPEL671,CAPEL672}, which is a generalization of the Ising model~\cite{Ising25}, is one of the widely studied models in statistical mechanics. It has been used to study the disruption of the $^\mathrm{4}$He superfluid transition by an admixture of $^\mathrm{3}$He~\cite{Blume71}, relaxation dynamics in molecular-based single-chain magnets~\cite{Kishine06}, hysteresis in FeRh films~\cite{Maat05}, and so on. This model has also attracted particular attention in connection with its wetting and interfacial adsorption under the presence or absence of bond randomness~\cite{Selke83,Selke84,Fytas13,Fytas19}. 
Considerable interest has centered recently on investigating dynamic phase transitions and randomness~\cite{Vasilopoulos21,Vatansever18,Vatansever18_2}. The study of the effects of random-anisotropy in the BC model has been proposed to describe the phase separation and the critical behavior of  $^3$He-$^4$He in random media~\cite{Maritan92,Buzano94}. The BC model consists of the pair-wise Ising interactions of spins, each of which can take on the values ``$-1$'', ``$+1$'', or ``$0$'' to allow for vacancies. The density of impurities, sites with ``$0$'' state, is controlled by the anisotropy field $\Delta$ (also called single-ion-splitting crystal field), since it gives rise to a zero-field splitting, raising the energy of the ``$\pm 1$'' states above the ``0'' state. The most interesting feature of this model is its unusual rich phase diagram involving a non-trivial tricritical point (TCP): a change in the nature of the phase transition from first-order to continuous. In one of the most recent studies~\cite{Azhari20}, it has been demonstrated numerically that the TCP is independent of $S$ in the mixed spin-$1/2$ and spin-$S$ BC model with $S = 1$, $2$, and $3$ in three-dimensional lattices. In a numerical study, the accurate location of TCP is troublesome and not easy to achieve in a multidimensional space of coupling constants. The principal difficulty in this model and related ones (e.g. the Blume-Emery-Griffiths model~\cite{Blume71} and the Potts model~\cite{Potts52}) is often encountered when dealing with the exceptionally large fluctuations in the order parameter close to the TCP~\cite{Beale86,Landau81}. 
 
The BC model has been studied extensively by various approaches, such as mean-field theory~\cite{Blume66,Capel66,CAPEL671,CAPEL672}, Monte Carlo (MC) simulations and MC renormalization group calculations~\cite{Berker76,Kaufman81,Selke83,Landau81,Landau86,Xavier98}, $\epsilon$-expansion renormalization  groups~\cite{STEPHEN73, Chang74,Tuthill75, WEGNER75}, high- and low-temperature series expansions~\cite{Burkhardt76,Camp75}, phenomenological finite-size scaling ansatz analysis~\cite{Nightingale82,Beale86}, and transfer-matrix method~\cite{Jung17}. Recently, notable progress has been achieved thanks in large part to the multicanonical simulations such as the Wang-Landau (WL) algorithm~\cite{Wang01}. Since the work of Silva et al.~\cite{Silva06}, the WL algorithm has been used extensively in the BC model to extract precise data of thermodynamic properties in both the first-order and continuous phase transition boundaries~\cite{Kwak15,Fytas11,Fytas13,Malakis09,Malakis10,Azhari20}. However, one should notice that these studies are mostly restricted to regular lattices, namely, the square, triangular, and simple-cubic lattices, with the exception of a few works on small-world networks without crystal field~\cite{Fernandes10,Lima13} and a fractal structure with a scale-free degree distribution~\cite{RochaNeto18}. In general, the role of lattice structure on the phase diagram of the BC model has been overlooked.

In this paper, we fill the gap by examining the BC model on three kinds of nonregular lattices: the Lieb lattice (LL), the decorated triangular lattice (DTL), and the decorated simple cubic lattice (DSC) (see figure~\ref{fig:fig1}). For the sake of generality, both two- and three-dimensional lattices are considered, which makes it straightforward to analyze the particular cases of the DTL and the DSC lattices because they share the same coordination number. We present the phase diagrams of the BC model for the three nonregular lattices. We locate the position of the TCP and also calculate the critical exponents for continuous phase transitions. For this purpose, we employ extensive MC simulations based on a combination of the Metropolis update (MU)~\cite{Metropolis53} and WL algorithm~\cite{Wang01}, which allows access to large system sizes and first-order phase transitions. The outline of this paper is the following. In Sec.~\ref{Sec:Method} we introduce the Hamiltonian of the BC model together with a discussion of the numerical scheme and the useful physical observables, necessary for the application of finite-size scaling (FSS) analysis. Section~\ref{Sec:Results} is devoted to the discussion of our main findings followed by a summary of our conclusions in Sec.~\ref{Sec:Conclusion}.

\section{Model and Methods \label{Sec:Method}}

\begin{figure}
\includegraphics[width=0.7\columnwidth]{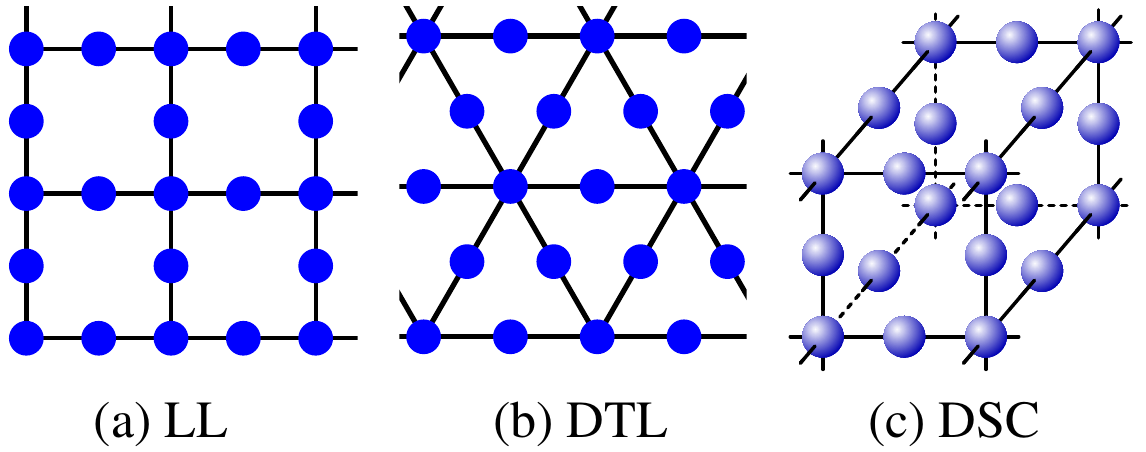}
\caption{\label{fig:fig1} 
(Color online) The three kinds of lattices studied in this paper: Lieb lattice (LL), decorated triangular lattice (DTL), and decorated simple cubic lattice (DSC).}
\end{figure}

We study the BC model characterized by the Hamiltonian
\begin{eqnarray}
H=-J \sum_{\langle i, j \rangle} S_i S_j + \Delta \sum_{i} S_i ^2 , \label{Hamiltonian1}
\end{eqnarray}
where spin $S_i$ may take on the values $\{-1, 0, 1\}$ and sits on site $i$ of one of the considered nonregular lattices. Periodic boundary conditions are employed. The parameters $J$ and $\Delta$ represent the ferromagnetic exchange interaction ($J>0$) and the single-ion-splitting crystal field, respectively.
In the limit $\Delta \rightarrow - \infty$, the vacancies are completely suppressed and the system behaves exactly as the conventional two-state Ising model. For regular lattices, the phase diagram of the BC model in the $\Delta/J$-$T/J$ plane is well known and fully understood. In $d\geq2$ dimensions, continuous and first-order phase transition lines separate the ferromagnetic (FM) phase from the paramagnetic (PM) phase. These lines meet at the TCP, which is characterized by $\Delta=\Delta_t$. The PM phase can be a random arrangement of spins at high temperature or a $\pm1$-spin gas dominated by nonmagnetic environment ($0$-spin) in case of high crystal fields and low temperatures. For $-\infty \le \Delta \le \Delta_t$, the BC model undergoes a continuous phase transition, which shares the same critical exponents with the Ising model, whereas for $\Delta_t \le \Delta \le \Delta_{\mathrm{crit}}$ the transition becomes first-order. No transition is observed for $\Delta > \Delta_{\mathrm{crit}}$. However, the effect of the non-regularity of the lattice can change the characteristic behavior of the model, as will be shown below. We analyze the phase diagrams in detail and check the criticality in three kinds of nonregular lattices.

In this work, we consider three kinds of two- and three-dimensional nonregular lattices: LL, DTL, and DSC. The number of lattice points is $N=BL^d$, where $L$ is the linear size of the system, $d$ is the spatial dimension, and $B$ is the number of sites per unit cell. For LL, $B=3$; for DTL and DSC, $B=4$.
In LL, 2/3 of sites have the coordination number $z_1=2$ and the other 1/3 of sites have $z_2=4$ to give average coordination number $Z=8/3$. As for DTL and DSC, 3/4 of sites have $z_1=2$ and the other 1/4 of sites have $z_2=6$, and so $Z=3$.
(See figure~\ref{fig:fig1}.)

For the simulation of the BC model on these lattices, we have employed the importance-sampling MC method using the local MU~\cite{Metropolis53} as well as the WL sampling~\cite{Wang01, Silva06, Fytas11, Kwak15}. Although the MU method is simple and easy to implement, it often has difficulties exploring the system close to first-order transitions because there is a possibility that the random walker gets trapped in a local minimum, which leads to supercritical slowing down effect~\cite{Janke94}. In the case of continuous transitions, the critical slowing down reduces significantly the efficiency of the algorithm near the critical temperature~\cite{Janke94}. 
In contrast, the WL sampling overcomes the critical and supercritical slowing down; it eliminates hysteresis and is straightforwardly applicable to the BC model. A further interesting fact that highlights the importance of the WL is that physical quantities, for any temperature and anisotropy, can be obtained just by one calculation. On the other hand, the MU allows us to set one single set of temperature and anisotropy per run, which prevents us to obtain physical observables as continuous functions. Nevertheless, it provides access to simulations on large lattice sizes, which makes it suitable for the FSS studies; but convergences of large lattice sizes are always difficult in the WL scheme since the Hamiltonian of the model has a multiparametric nature and hence a huge number of energy levels. The maximum sizes we used in this work within the WL algorithm are $N=768$ ($L=16$), $N=576$ ($L=12$), and $N=864$ ($L=6$) for LL, DTL, and DSC, respectively.
Since the Metropolis implementation has been widely discussed in literatures (see, e.g., \cite{Newman99}), we shall present only calculation details here. At each temperature, we performed $1\times10^8$ MC steps to compute the average value of the physical quantities after discarding $5\times10^6$ MC steps to ensure equilibrium. One Monte Carlo step consists of $N$ attempts of spin flip by the Metropolis algorithm. We used 24 nodes within the message passing interface (MPI) incorporated with the Mersenne Twister pseudorandom number generator for distributed parallel environments~\cite{Matsumoto00}. The whole calculation was repeated at least two times: one time with increasing temperature and the other time with decreasing temperature. No hysteresis is observed except close to the first-order phase transition in the reentrant region. Near the continuous transition, more iterations were performed to overcome critical slowing down and to get reliable results.
The WL algorithm directly calculates the density of states $\rho(E_1, E_2)$ via a random walk in energy space with the transition probability
\begin{eqnarray}
P[(i_1, i_2)\rightarrow (j_1, j_2)] = 
   \mbox{min}\left[1, \frac{\rho(E_{i_1}, E_{i_2})}{\rho(E_{j_1}, E_{j_2})}\right],
\end{eqnarray}
which makes the histogram $h(E_1, E_2)$ flat.
The two energy variables $E_1$ and $E_2$ represent the two terms of the Hamiltonian in (\ref{Hamiltonian1}), respectively:
\begin{eqnarray}
E_1 = \sum_{\langle i, j \rangle} S_i S_j ~ 
\mbox{ and } ~
E_2 = \sum_{i} S_i^2 .
\end{eqnarray}
At each step, the WL refinement is $\rho(E_{1}, E_{2}) \rightarrow f_n \, \rho(E_{1}, E_{2})$, where $f_n>1$ is an empirical factor. Whenever the energy histogram is flat enough, the modification factor $f_n$ is adjusted as $f_{n+1}= \sqrt{f_n}$ with $f_0 = e$ and a new set of random walks is performed. The whole simulation is terminated when $f_n$ becomes close enough to 1: $f_\mathrm{final} < \exp(10^{-10})$. See \cite{Yu15,Azhari20} for more details. During the simulation, average values of thermodynamic observables $O(E_{1}, E_{2})$ as a function of $E_1$ and $E_2$ should be calculated.
Once the density of states $\rho(E_{1}, E_{2})$ is obtained, the partition function can be calculated for any value of temperature and anisotropy,
\begin{eqnarray}
\mathfrak{Z}(T,\Delta)=\sum_{E_1, E_2} \rho(E_1, E_2) e^{-KH}, \label{eq:4}
\end{eqnarray}
where $K$ denotes the inverse temperature $1/k_B T$ and $k_B$ is the Boltzmann constant. 
From (\ref{Hamiltonian1}), $H=(-J E_1 + \Delta E_2 )$.
It is straightforward that all thermodynamic observables $\langle O \rangle(T,\Delta)$ can be calculated without additional simulation for each temperature and anisotropy:
\begin{eqnarray}
\langle O \rangle (T,\Delta)= \frac{1}{\mathfrak{Z}} \sum_{E_1, E_2} O(E_{1}, E_{2}) \rho(E_1, E_2)
e^{-K H} .
\end{eqnarray}
We use this canonical average to calculate magnetization and susceptibility:
\begin{eqnarray} 
&& m(T,\Delta) = \langle M \rangle , \label{eq:m3} \\
&&\chi(T,\Delta) = N K \left(\langle M^2 \rangle - \langle M \rangle^2 \right). \label{eq:m5}
\end{eqnarray}
The absolute magnetization per site $M$ is defined as $M=|\sum_i{S_i}|/N$.
We also investigate the Binder cumulant \cite{Binder81}
\begin{eqnarray}
U= 1 - \frac{\left\langle M^4 \right\rangle}{3 \left\langle M^2 \right\rangle ^2}
\end{eqnarray}
to locate the critical temperature $T_c$ and to determine the type of transition.
For both algorithms, the whole calculation was repeated at least two times using different pseudorandom number sequences to make an average and to estimate the statistical error. We confirmed that the statistical error is smaller than errors from finite-size lattices.

\begin{figure}
\includegraphics[width=0.8\columnwidth]{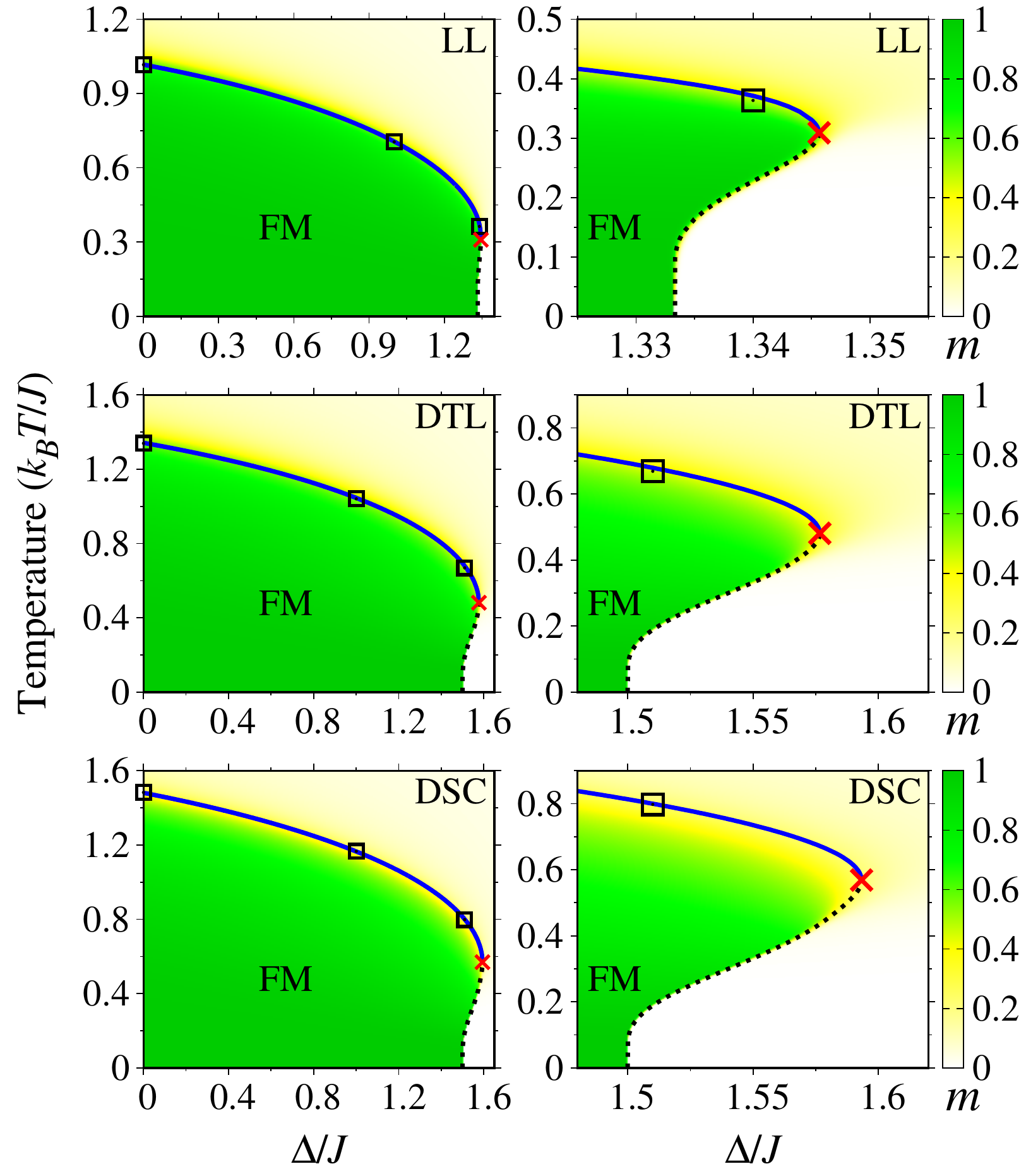}
\caption{\label{fig:fig2} 
(Color online) Phase diagrams of the Blume-Capel model in the Lieb lattice (LL), decorated triangular lattice (DTL), and decorated simple cubic lattice (DSC) obtained by the Wang-Landau algorithm. Solid lines denote continuous phase transitions and dotted lines denote first-order phase transitions. The location of the tricritical point is represented by a cross symbol (`$\times$'). The background color represents magnetization ($m$) in the lattices of $L=16$, $L=12$, and $L=6$ for the LL, DTL, and DSC, respectively. Values of the transition temperature calculated by the Metropolis algorithm are also represented by squares for comparison. The right panels magnify the area near the tricritical point and reentrance. The statistical error is smaller than the symbol size.}
\end{figure}

\begin{figure}
\includegraphics[width=\columnwidth]{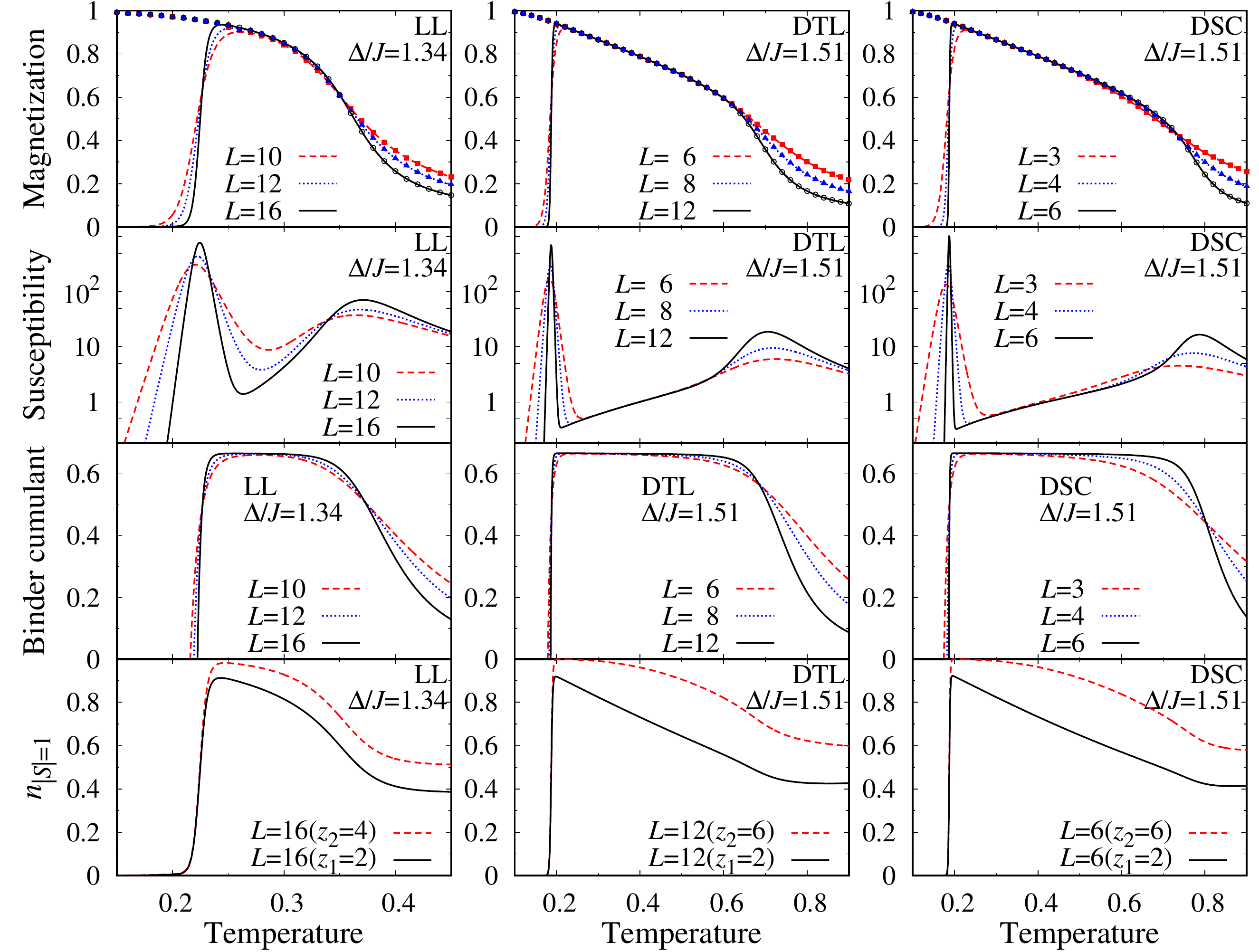}
\caption{\label{fig:fig3} 
(Color online) Magnetization, magnetic susceptibility, Binder cumulant, and fraction of sites with $S_i=\pm1$ as a function of temperature ($k_B T/J$) in the Lieb lattice (LL), decorated triangular lattice (DTL), and decorated simple cubic lattice (DSC). The values of anisotropy ($\Delta$) are between $\Delta_{\mathrm{crit}}$ and $\Delta_{t}$. 
In the lowest panels, $z_i$ denotes the coordination number of each sublattice.
Lines are obtained by the Wang-Landau algorithm; symbols in the upper panels represent magnetization calculated by the Metropolis algorithm as temperature is lowered. The statistical error is smaller than the symbol size.
}
\end{figure}

\begin{figure}
\includegraphics[width=0.7\columnwidth]{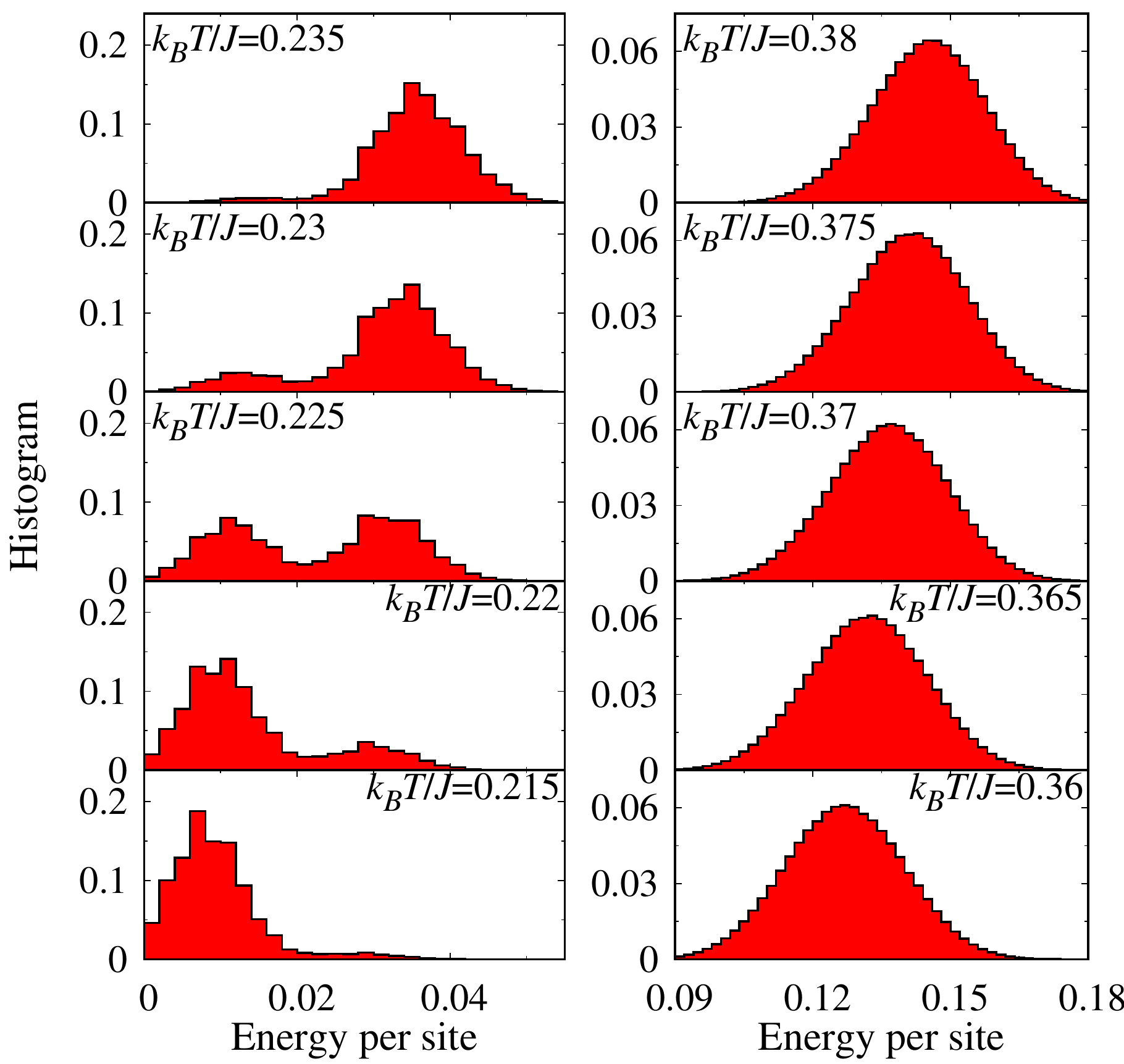}
\caption{\label{fig:fig4} 
Histogram as a function of energy per site for the Lieb lattice with $\Delta/J=1.34$ and $L=16$ near the first-order phase transition temperature $T_{c1}$ (left column) and continuous phase transition temperature $T_{c2}$ (right column) obtained by the Wang-Landau algorithm.}
\end{figure}

\begin{figure}
\includegraphics[width=0.7\columnwidth]{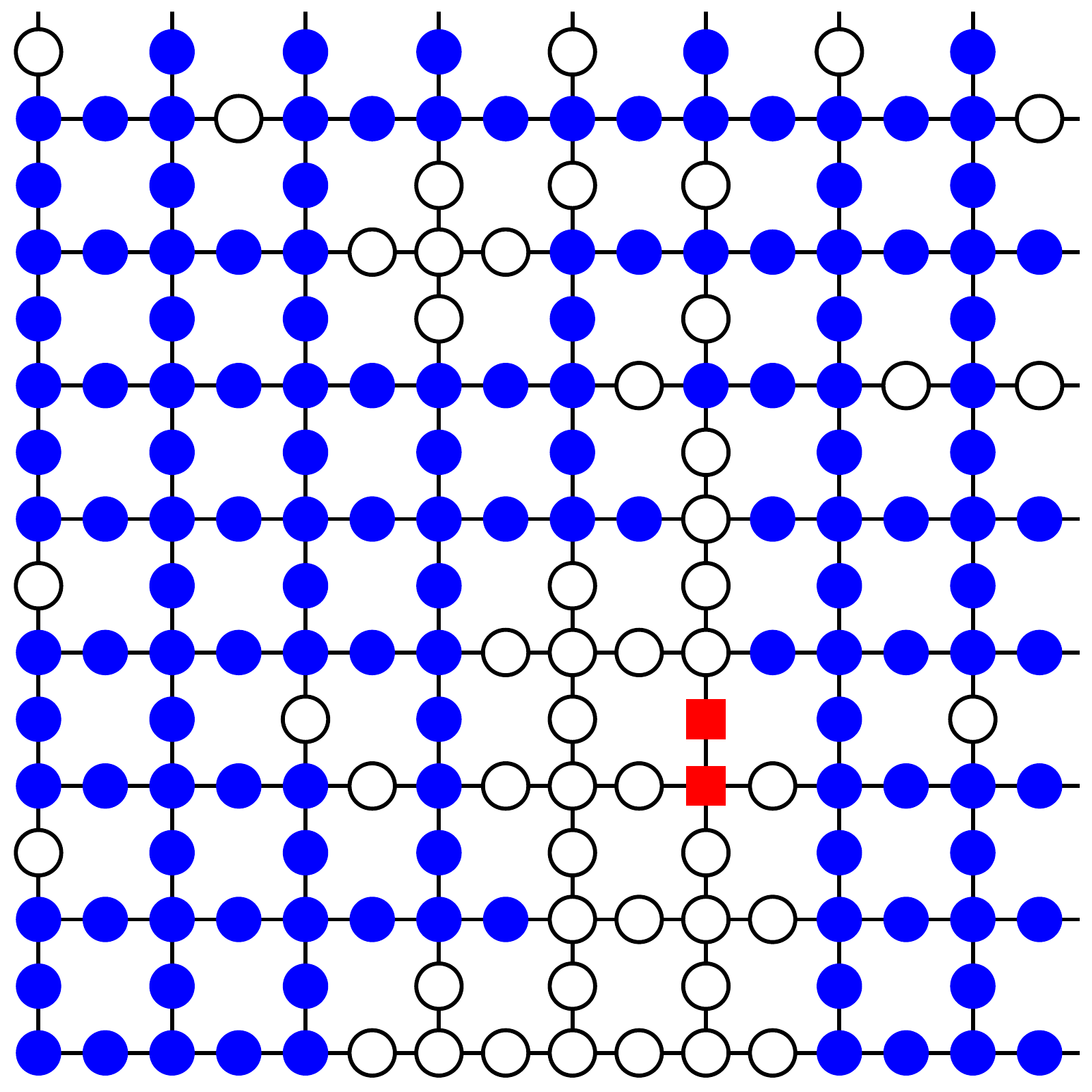}
\caption{\label{fig:fig5} 
(Color online) Typical spin configuration of the Blume-Capel model in the Lieb lattice at $\Delta/J=1.34$ and $k_B T/J = 0.3$. Blue solid circles, red solid squares, and empty circles represent values of spin $+1$, $-1$, and $0$, respectively, at each site.}
\end{figure}

\begin{figure}
\includegraphics[width=\columnwidth]{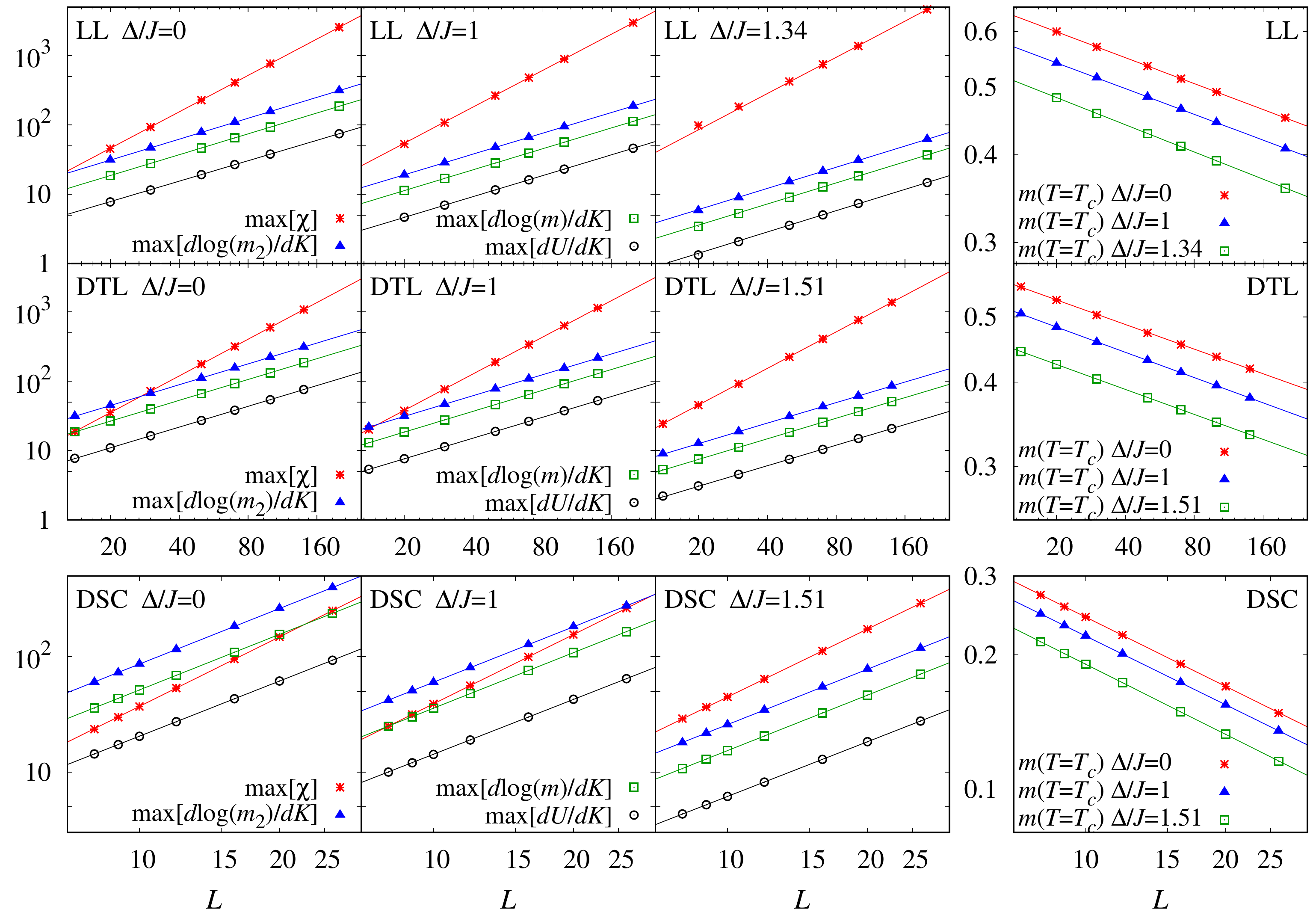}
\caption{\label{fig:fig6} 
(Color online) Maximum values of magnetic susceptibility ($\chi$), $d\log(m_2)/dK$, $d\log(m)/dK$, and $dU/dK$, and the value of $m$ at $T_c$ as a function of linear lattice size $L$, where $m_2$, $m$, $U$, and $K$ represent square magnetization ($m_2=\langle M^2 \rangle$), magnetization, Binder cumulant, and inverse temperature ($K = 1/k_B T$), respectively.
These results were obtained by the Metropolis algorithm and the errorbars are within the symbol size. Straight lines are from fitting to  (\protect\ref{fsss11})-(\protect\ref{fsss15}).
}
\end{figure}

\section{Results and discussion \label{Sec:Results}}
In this section, we present our numerical results regarding the BC model on nonregular lattices, for which some interesting parts of the phase diagram are discussed based on our analysis of physical observables mentioned above. We compare our findings with those of the regular lattices. 
The issues about critical exponents and universality class for continuous transitions are also discussed.

\subsection{Phase diagrams \label{Sec:A}}

We determined the behavior of the transition temperature $T_c$ as a function of the crystal field $\Delta$, as shown in figure~\ref{fig:fig2}, covering both first-order and continuous transition regimes of the phase diagrams of the LL, DTL, and DSC. The transition temperature is obtained by locating the size-independent crossing point of the Binder cumulants for different lattice sizes. This method can be used for first-order as well as continuous phase transitions~\cite{Challa86,Cary17,Azhari20}. Since WL density of states gives the thermodynamic quantities for continuous values of $T$ and $\Delta$, $T_c$ can be obtained continuously as a function of $\Delta$; the solid and dotted lines in figure~\ref{fig:fig2} represent continuous and first-order phase transitions, respectively. The location of the TCP is indicated by a cross symbol (`$\times$'). The background color in figure~\ref{fig:fig2} depicts the temperature dependence of the magnetization $m$ of the BC model in a finite lattice: deep green color indicates ferromagnetic ordered state and bright white color represents completely disordered state or zero-spin state. Squares denote the MU results for much larger lattices ($N=120\,000$ ($L=200$) for LL), which illustrate the excellent agreement between the two methods. It shows that the error by the correction to scaling~\cite{Ferrenberg18}, if it exists, is small.

Overall, the phase diagrams of the BC model on nonregular lattices in figure~\ref{fig:fig2} are qualitatively different from those on regular peers: the transition line of the BC model on regular lattices decreases monotonically as $\Delta$ increases until $T_c=0$ and vanishes at $\Delta=\Delta_{\mathrm{crit}}$. On the contrary, for nonregular lattices, the critical transition line extends up to $\Delta=\Delta_{t}$ where $\Delta_{t}>\Delta_{\mathrm{crit}}$, while the first-order transition line connects the TCP and the quantum critical point $(T=0,\Delta=\Delta_{\mathrm{crit}})$, which manifests as a reentrance phase transition in the region $\Delta_{\mathrm{crit}}<\Delta<\Delta_{t}$. Figure~\ref{fig:fig3} shows physical quantities for different system sizes as a function of temperature for $\Delta$ between $\Delta_{\mathrm{crit}}$ and $\Delta_{t}$: $\Delta/J = 1.34$ for LL, $\Delta/J = 1.51$ for DTL, and $\Delta/J = 1.34$ for DSC. Spontaneous magnetization appears for temperature between the two transition temperatures. Usually, the type of the transition is indicated by the slope of the order parameter; the magnetization data close to the lower and higher temperatures $T_{c1}$ and $T_{c2}$ respectively imply the first-order and continuous character of the transitions. The lines in figure~\ref{fig:fig3} are obtained from the WL algorithm, while the symbols are given by the MU method as temperature is lowered. The disagreement between the two results at low temperatures is associated with the dynamics in the canonical ensemble, which suffers from the ``supercritical slowing down'' near the first-order transition. For small system sizes, this can be cured by increasing the number of MC steps and changing the temperature-process (comparisons of hot and cold starts or even mixed configurations)~\cite{Azhari20}, an optimization of the algorithm is needed for rather big lattice sizes \cite{Martin-Mayor07}. The second and third panels of figure~\ref{fig:fig3} give further validations of the nature of the two consecutive transitions. In the vicinity of $T_{c1}$, susceptibility data show sharp peaks with little size dependence and the Binder cumulant has a negative valley, which indicate a first-order phase transition in the thermodynamic limit~\cite{Binder84,Tsai98,Crokidakis10,Cary17}. Whereas such behaviors do not appear near $T_{c2}$, which strongly supports the continuous transition. Further proof is realized in the energy histogram close to these two transitions~\cite{Cary17} as presented in figure~\ref{fig:fig4} for LL with $N=768$ ($L=16$); the typical single- and double-peak structures confirm the continuous and first-order transitions, respectively. 
We confirmed that the transition at higher temperature $T_{c2}$ is continuous while the transition at lower temperature $T_{c1}$ is first-order in the whole reentrant regime. Therefore, the TCP exists where the first-order transition line ($T_{c1}$) and the continuous transition line ($T_{c2}$) meet and the reentrance disappears.
Our estimation of the TCPs in the phase diagram are ($\Delta_t/J=1.3457(1)$; $k_BT_t/J=0.309(2)$), ($\Delta_t/J=1.5766(1)$; $k_BT_t/J=0.481(2)$), and ($\Delta_t/J=1.5933(1)$; $k_B T_t/J=0.569(4)$) for LL, DTL, and DSC, respectively.

Since the reentrance is incurred naturally by frustration or randomness in many lattice systems (see \cite{Creighton11,Diep13,Yusuf12} and references therein), it is intriguing to observe such a behavior in unfrustrated system like this. To figure out the mechanism and nature of the ordering in the reentrance area, we investigated the density of non-zero spin $n_{\left|S\right|=1}$ as shown in the lower panels of figure~\ref{fig:fig3}. At low temperature, the zero-spin (ZS) state is dominant up to $T_{c1}$, then the non-zero spin density $n_{\left|S\right|=1}$ increases abruptly to induce the FM phase. As the temperature increases further above $T_{c2}$, $n_{\left|S\right|=1}$ decreases and converges to 2/3. Therefore, we conclude that the transition at $T_{c1}$ is from ZS phase to FM phase, while the transition at $T_{c2}$ is from FM phase to PM phase.   Interestingly, $n_{\left|S\right|=1}$ depends on the local environment (for example, $z_1=2$ and $z_2=4$ in LL) as long as $T>T_{c1}$. In this case, $n_{\left|S\right|=1}$ is close to 1 for sites with larger coordination number. 

To understand the reentrant behavior, we examine this model close to the quantum critical point at $(T=0, \Delta=\Delta_{\mathrm{crit}})$, where the complete FM and the complete ZS states coexist and then the ground state is degenerate. The energies of these states are $E_{\mathrm{FM}}=-JNZ/2 + N\Delta$ and $E_{\mathrm{ZS}}=0$, respectively, where $Z$ is the average coordination number. Moreover, $E_{\mathrm{FM}}=E_{\mathrm{ZS}}$ implies $\Delta_{\mathrm{crit}}=JZ/2$. As we switch on the temperature, we create spin-flip excitations in the system. Let us consider the lowest spin-flip excitations (one zero-spin-flip for complete FM and one up-spin-flip for complete ZS) and abbreviate the resulting states as FM$'$ and ZS$'$, respectively. 
Since the entropies of the two states FM$’$ and ZS$’$ have the same order of magnitude, we have only to compare their respective energies. However, there are different sublattices in nonregular lattices. Let us define the coordination number of each sublattice by $z_i$. Such lattices have at least one of $z_i$ less than the average coordination number $Z$. The two excited states have the energies $E_{\mathrm{FM}’}=J z_i-\Delta_{\mathrm{crit}}$ and $E_{\mathrm{ZS}’}=\Delta_{\mathrm{crit}}$, where $z_i$ is the coordination number of the flipped site. The difference $E_{\mathrm{FM}’}- E_{\mathrm{ZS}’}$ gives $J(z_i-Z)$, which is negative for $z_i$ less than the average coordination number $Z$. Therefore, a FM$'$ state has a lower energy than a ZS$'$ state if the flipping occurs at a site with coordination number less than $Z$. In other words, at low temperature and for $\Delta=\Delta_{\mathrm{crit}}$, the low-lying excited states favor FM ordering with zero-spin excitations on less-coordinated sites, which makes the phase boundary $T_{c1}$ between FM and ZS states  bend toward large values of $\Delta$, and then gives rise to a reentrant behavior in the region $\Delta_{\mathrm{crit}}<\Delta<\Delta_t$. This is consistent with the fact that $n_{\left| S \right|=1}$ is smaller in less-coordinated sites as shown in figure~\ref{fig:fig3}. In this case, the reentrance is due to the non-regularity of the lattice.

Figure~\ref{fig:fig5} illustrates a typical instant spin configuration for LL in the FM state between the consecutive transitions at $k_B T/J = 0.3$ and for $\Delta/J=1.34$. In the background of the dominating FM order, some less-coordinated sites have zero-spin state. These zero-spin sites may cooperate to induce clusters of zero-spin sites. Within the zero-spin clusters, opposite spin clusters rarely appear. As temperature increases through the phase transition $T_{c2}$, zero-spin clusters grow gradually to break long-range ordering of the FM state. Although only the case of LL is shown, there is no qualitative difference in the other nonregular lattices considered in this work.

\begin{table}
\caption{Critical exponents and critical Binder cumulant $U^*$ for the Blume-Capel model in the Lieb lattice (LL), decorated triangular lattice (DTL), and decorated simple cubic lattice (DSC). 
Reference values for the Ising model in two dimensions (2D)~\protect\cite{Onsager,Kamieniarz93,Selke07,Selke09} and in three dimensions (3D)~\protect\cite{Kos16,Deng03,Yu15PhA} are also presented. In the rightmost column, SQL, TL, and SCL abbreviate square lattice, triangular lattice, and simple cubic lattice, respectively.}
\label{table:critical_exponents}
\begin{tabular}{c c c c c c} 
\br
  & $\Delta/J$ & $\nu$ & $\gamma/\nu$ & $\beta/\nu$ & $U^*$ \\
\mr
 LL & 0    & 1.01(1) & 1.75(1) & 0.123(4) & 0.612(2)\\ 
 LL & 1    & 1.00(1) & 1.74(1) & 0.122(6) & 0.612(2)\\ 
 LL & 1.34 & 0.99(1) & 1.74(1) & 0.129(6) & 0.611(2)\\
\mr
 DTL & 0    & 1.00(1) & 1.76(1) & 0.121(4) & 0.614(2)\\
 DTL & 1    & 1.01(1) & 1.75(1) & 0.125(4) & 0.613(2)\\
 DTL & 1.51 & 1.01(1) & 1.75(1) & 0.123(4) & 0.613(2)\\
\mr
 DSC & 0    & 0.627(3) & 1.99(2) & 0.516(6) & 0.463(3)\\
 DSC & 1    & 0.626(4) & 1.98(2) & 0.511(6) & 0.466(3)\\
 DSC & 1.51 & 0.629(4) & 1.96(2) & 0.521(6) & 0.463(3)\\
\mr
 \multirow{2}{*}{2D-Ising} & & \multirow{2}{*}{1} & \multirow{2}{*}{1.75} & \multirow{2}{*}{0.125} & 0.6106901(5) (SQL) \\
                           & &                    &                       &                        & 0.61182864(2) (TL) \\ 
 3D-Ising & & 0.629971(4) & 1.963702(2) & 0.5181489(10) & 0.46531(4) (SCL)  \\
 \br
\end{tabular}
\end{table}

\begin{figure}
\includegraphics[width=0.7\columnwidth]{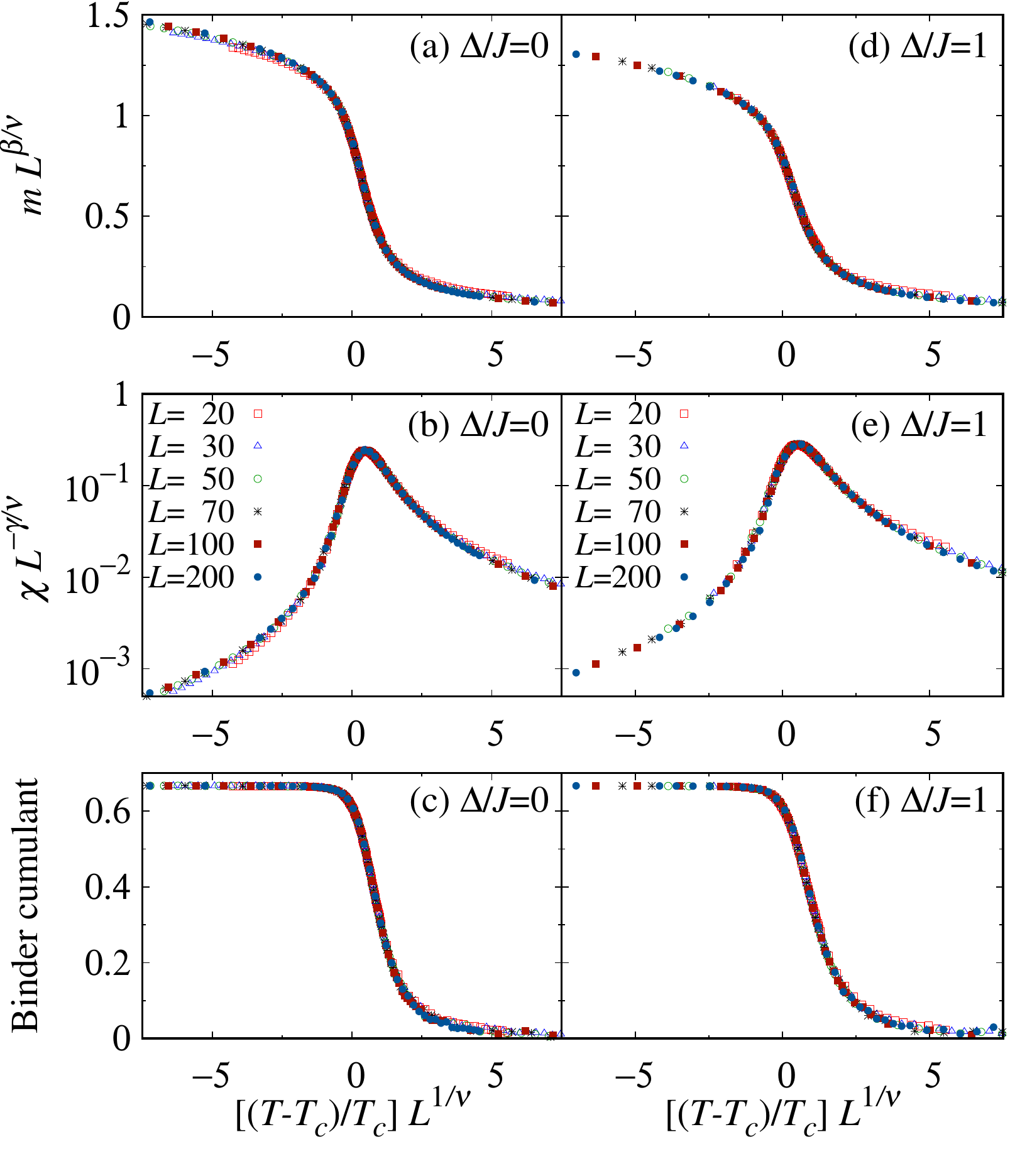}
\caption{\label{fig:fig7} 
(Color online) Magnetization, susceptibility, and Binder cumulant of the Blume-Capel model in the Lieb lattice as a function of rescaled temperature. These results were obtained by the Metropolis algorithm. The errorbars are within the symbol size.}
\end{figure}

\subsection{Critical exponents of the continuous transitions in the Blume-Capel model}

Motivated by the rich phase diagrams of Sec.~\ref{Sec:A} and to gain further insight into the critical behavior of the BC model on nonregular lattices, its singularities around the continuous transitions are analyzed. For each lattice, we have calculated the critical exponents for three different values of the anisotropy: $\Delta/J \in \big\{ 0, 1, 1.34 \textrm{ (for LL)}, 1.51 \textrm{ (for DTL and DSC)} \big\}$. We calculated the three exponents $\gamma$, $\nu$ and $\beta$ using FSS~\cite{Fisher72,Brezin85,Ferrenberg91}. In a finite system of linear size $L$, it is well known that the physical quantities near the critical point scale as
\begin{eqnarray}
&&m(L,x) = L^{-\beta/\nu} \Psi_{m}(x)(1+\ldots), \label{fsss1} \\
&&\chi(L,x) = L^{\gamma/\nu} \Psi_{\chi}(x)(1+\ldots), \label{fsss4}\\
&&U(L,x) = \Psi_{U}(x)(1+\ldots) , \label{fsss5}
\end{eqnarray}
where $\Psi$ characters represent universal scaling functions with argument $x=[(T-T_{c})/T_c]L^{1/\nu}$. The dots in $(1+\ldots)$ stand for the corrections to the scaling behavior; they are less than statistical error in our calculations and ignored. Equations~(\ref{fsss1})-(\ref{fsss5}) are used to derive the following FSS relations:
\begin{eqnarray}
&&\mbox{max}[\chi] \propto  L^{\gamma/\nu}, \label{fsss11}\\
&&\mbox{max}\left[\frac{\rmd\log(m)}{\rmd K}\right] \propto  L^{1/\nu} , \label{fsss12}\\
&&\mbox{max}\left[\frac{\rmd\log(m_2)}{\rmd K}\right] \propto  L^{1/\nu} , \label{fsss13}\\
&&\mbox{max}\left[\frac{\rmd U}{\rmd K}\right] \propto  L^{1/\nu} , \label{fsss14}\\
&&m(T=T_c) \propto L^{-\beta/\nu} ,\label{fsss15}
\end{eqnarray}
where $m_2=\langle M^2 \rangle$.

The results of the FSS analysis for different values of the crystal field $\Delta$, selected on both sides of the critical value $\Delta_{\mathrm{crit}}$, are collected in table~\ref{table:critical_exponents} along with the known reference values of critical exponents for the Ising model in two dimensions~\cite{Onsager} and in three dimensions~\cite{Kos16}. These critical exponents are obtained by fitting the curves in figure~\ref{fig:fig6} to (\ref{fsss11})-(\ref{fsss15}), which are consistent with the continuous phase transition values for the Ising model~\cite{Onsager,Kos16}. Figure~\ref{fig:fig7} confirms that the FSS works quite well to make the measurement data at different lattice sizes $L$ collapse on the scaling functions $\Psi$.
As shown in table~\ref{table:critical_exponents}, the values of critical Binder cumulant $U^*$, which is the Binder cumulant at $T_c$ extrapolated to the thermodynamic limit, indicate the universality class as well. It is well known that $U^*$ depends on the boundary conditions and the shape of the lattice, but is independent of spin value or the lattice structure~\cite{Nicolaides88,Kamieniarz93,Selke07,Selke09,Yu15PhA}. The values of $U^*$ for LL and DTL are independent of $\Delta$ and consistent with those of the Ising model on the square lattice ($U^*=0.6106901(5)$) and on the triangular lattice ($U^*=0.61182864(2)$), respectively, within error bars~\cite{Kamieniarz93,Selke07,Selke09}. For DSC, $U^*$ is also consistent with the Ising model on the simple cubic lattice ($U^*=0.46531(4)$)~\cite{Deng03,Yu15PhA}.
The results of critical exponents and critical Binder cumulant clearly indicate that the BC model on nonregular lattices, in the continuous phase transition regime, belongs to the universality class of the Ising model.

\section{Conclusions \label{Sec:Conclusion}}

In this study, we presented a numerical study of the BC model on nonregular lattices: LL, DTL, and DSC. By means of the Metropolis and the WL algorithms and supplemented by FSS analysis, we have carefully analyzed the phase diagrams, the critical behavior, and the location of the TCP. We found that the nonregular character of lattices induces the reentrance behavior in the region $\Delta_{\mathrm{crit}} < \Delta < \Delta_{t}$ of the phase diagrams. Furthermore, we proposed in Sec.~\ref{Sec:A} a physical mechanism for the reentrant behavior based on our investigation of the density of non-zero spin $n_{\left|S\right|=1}$ and the stability analysis of spin configurations at low temperature and close to the quantum critical point $\Delta=\Delta_{\mathrm{crit}}$. It turns out that the BC model always has a double transition as long as the lattice considered is nonregular, and the reentrance is more pronounced if the difference between $Z$ and the smallest $z_i$ is large. On the other hand, we located the TCP with very high precision as ($\Delta_t/J=1.3457(1)$; $k_B T_t/J=0.309(2)$), ($\Delta_t/J=1.5766(1)$; $k_B T_t/J=0.481(2)$), and ($\Delta_t/J=1.5933(1)$; $k_B T_t/J=0.569(4)$) for the LL, DTL and DSC, respectively. Finally, we checked the universality class for three different values of the crystal field: $\Delta/J \in \big\{ 0, 1, 1.34 \textrm{ (for LL)}, 1.51 \textrm{ (for DTL and DSC)} \big\}$. The critical exponents and the critical Binder cumulant are found to be consistent with the reference values of the Ising model in two and three dimensions.

\section*{Acknowledgments}
This work was supported by the National Research Foundation of Korea(NRF) grant funded by the Korea government(MSIT) (No. 2021R1F1A1052117). M. A. gratefully acknowledges support by the German Research Council (DFG) via the Research Unit FOR 2316. M. A. would like to thank Andreas Kl\"umper and Rachid A\"it Mouss for interesting discussions and helpful comments.

\section*{References}
\bibliography{Ref}

\end{document}